\documentstyle[12pt]{article}
\newcommand{\be}{\begin{equation}}
\newcommand{\en}{\end{equation}}
\newcommand{\bea}{\begin{eqnarray}}
\newcommand{\ena}{\end{eqnarray}}
\newcommand{\Det}{\hbox{Det}}
\newcommand{\hbo}{\hbox to 1 true cm {\hfill } }
\newcommand{\tr}{\hbox{tr}}
\newcommand{\Tr}{\hbox{Tr}}

\def\dslash{\partial\kern-.5em\slash}
\def\kslash{k\kern-.5em\slash}
\def\Chi{{\mathop{\kern 2pt\vcenter{\hbox{$\chi $}}\kern2pt}}}
\begin{document}
\baselineskip=16truept
\vglue 1truecm

\vbox{
\hfill  November 5, 1992
}

\vfil
\centerline{\bf \large Instantons in field strength formulated
   Yang-Mills theories$^1$}

\bigskip
\centerline{H.\ Reinhardt, K.\ Langfeld, L.\ v.\ Smekal }
\medskip
\centerline{Institut f\"ur theoretische Physik, Universit\"at
   T\"ubingen}
\centerline{D--7400 T\"ubingen, Germany }
\bigskip

\vfil
\begin{abstract}

It is shown that the field strength formulated Yang-Mills theory
yields the same semiclassics as the standard formulation in terms
of the gauge potential. This concerns the classical instanton
solutions as well as the quantum fluctuations around
the instanton.

\end{abstract}

\vfil
\hrule width 5truecm
\vskip .2truecm
\begin{quote}
$^1$ Supported by DFG under contract Re $856/1 \, - \, 1$
\end{quote}
\eject
%
%
\leftline{\it 1.\ Introduction \/ }
Yang-Mills (YM) theories can be reformulated entirely
in terms of field strength thereby eliminating the gauge
potential~\cite{ha77,sch90,re90a}. The field strength formulation allows
for a non-perturbative treatment of the gluon self-interaction
and offers a simple description of the non-perturbative vacuum
where the gauge bosons are presumable condensed. In the so called FSA an
effective action for the field strength is obtained which
contains a term of order $\hbar $ with an explicit energy scale.
Hence in the FSA the anomalous breaking of scale invariance
is described already at tree level~\cite{sch90}. At this level
the YM vacuum is determined by homogeneous field configurations
with constant $\langle F^{a}_{\mu \nu } F^{a}_{\mu \nu } \rangle \not= 0$
and instantons do no longer exist as stationary points of the
effective action. This is already dictated by the fact that the effective
action contains an energy scale. This energy scale cannot tolerate
instantons which have a free scale (size) parameter.

In this paper we show that if the FSA is consistently formulated in
powers of $\hbar $ one recovers the same semiclassics as in the
standard formulation of YM theory in terms of the gauge potential.
All instantons which extremize the standard YM action are also
stationary points of the action to $O(\hbar )$ of the FSA.
Furthermore the leading $\hbar $ corrections originating from the
integral over quantum fluctuations around the instanton are
also the same in both approaches. Although one might have expected
this result on general grounds it is completely non-trivial
how this result emerges in the field strength formulation.
Furthermore the equivalence proof will also shed some new light
on the FSA.

\bigskip
\leftline{\it 2.\ Semiclassical approximation to the
Yang-Mills theory \/ }
We start from the generating functional of Euclidean YM theory
\be
Z[j] \; = \; \int {\cal D} A \;
\delta \bigl( f^{a}(A) \bigr) \, \Det {\cal M}_{f} \;
\exp \{ - S_{YM} [A] \, + \, \int d^{4}x \;
j A \}  \; ,
\label{eq:1}
\en
where
\be
S_{YM} \; = \; \frac{1}{4 g^{2} } \int d^{4}x \;
F^{a}_{\mu \nu }(A) F^{a}_{\mu \nu }(A)
\; , \hbo
F^{a}_{\mu \nu }(A) \; = \; \partial _{\mu } A^{a}_{\nu } -
\partial _{\nu } A^{a}_{\mu } +  f^{abc} A^{b}_{\mu } A^{c}_{\nu } \; ,
\label{eq:1a}
\en
is the classical YM action.
Furthermore $\delta \bigl( f^{a}(A) \bigr)$ is the gauge fixing constraint
and $ \Det {\cal M}_{f} $ denotes the Faddeev-Popov determinant.

It is well known that all finite action self-dual or anti self-dual
field configurations extremize the YM action~\cite{inst} i.e.\
solve the classical YM equation of motion
\be
\partial _{\mu } F^{a}_{\mu \nu } \; = \;  - f^{abc} A^{b}_{\mu }
F^{c}_{\mu \nu } \; .
\label{eq:3}
\en
These classical solutions are referred to as instantons. Expanding
the fluctuating gauge field  around the classical instanton
solution $A^{inst}_{\mu }$ up to second order in the
quantum fluctuations and performing the integral in
semiclassical approximation one finds
\be
Z[j=0] \; = \; Q \;
( \Det ' {\cal D} ^{-1}_{YM} (x_{1},x_{2}) )^{-1/2}
\; e^{ - S_{YM} [A^{inst}] } \; ,
\label{eq:7}
\en
where
\be
{\cal D}^{-1}_{YM} (x_{1},x_{2}) ^{ab}_{\mu \nu } \; = \;
\frac{ \delta ^{2} S_{YM} [A] }{ \delta A^{a}_{\mu } (x_{1})
\delta A^{b}_{\nu } (x_{2}) } \mid _{ A_{\mu } = A_{\mu }^{inst} }
\en
and the prime indicates that the zero modes of ${\cal D} ^{-1}_{YM} $
have to be excluded from the determinant. This can be done
in the standard fashion and yields the factor $Q$ in (\ref{eq:7}).
Furthermore even when the zero modes are
excluded the determinant is still singular and needs regularization.
As will become clear later, for our purpose a regularization
scheme that only depends on eigenvalues is convenient, e.g.\
Schwinger's proper time regularization or $\zeta $-function
regularization.
The second variation of the YM action reads
\be
g^{2} {\cal D}^{-1}_{YM} (x_{1},x_{2}) ^{ab}_{\mu \nu } \; = \;
\hat{F} ^{ab}_{\mu \nu } (x_{1}) \, \delta (x_{1},x_{2})
\; + \; \frac{1}{2} \int d^{4}x \;
\frac{ \delta F^{c}_{\kappa \lambda }(x) }
    { \delta A^{a}_{\mu }(x_{1}) }
\frac{ \delta F^{c}_{\kappa \lambda }(x) }
    { \delta A^{b}_{\nu }(x_{2}) }  \; ,
\label{eq:10}
\en
where
\be
\hat{F} ^{ab}_{\mu \nu } \; = \; f^{abc} F^{c}_{\mu \nu }
\en
denotes the field strength in the adjoint representation and
\be
\frac{ \delta F^{c}_{\kappa \lambda }(x) }
    { \delta A^{a}_{\mu }(x_{1}) } \; = \;
[ \hat{D} ^{ca}_{\kappa }(x) \delta _{\mu \lambda } \, - \,
\hat{D} ^{ca}_{\lambda }(x) \delta _{\mu \kappa } ] \;
\delta (x- x_{1}) \; ,
\label{eq:13}
\en
with
\be
\hat{D}^{ab}_{\mu }(x) \; = \; \partial _{\mu } \delta ^{ab}
\; - \; \hat{A}^{ab}_{\mu } (x)
\label{eq:6}
\en
being the covariant derivative.
The functional determinant $\Det {\cal D}^{-1}_{YM} $ in an instanton
background has been explicitly evaluated by t'Hooft~\cite{tho76}.

\bigskip
\leftline{\it 3. Field strength formulated Yang-Mills theory \/}
Non-Abelian YM theory can be equivalently formulated in terms
of field strength \cite{ha77,sch90,re90a}. Inserting the
identity
\be
\exp\{ -{1\over 4g^2} \int d^4\!x \, F^2(A)\}= \int {\cal
D}\!\Chi \, \exp\left(-\int d^4\!x \{ {1\over 4}\Chi^a_{\mu\nu}
\Chi^a_{\mu\nu}
+ {i\over 2g} \Chi^a_{\mu\nu} F^a_{\mu\nu}(A) \} \right)
\en
into (\ref{eq:1}) we obtain
\be
Z \,=\, \int {\cal D}\!\Chi  \; {\cal D}\!A \; \delta ( f^a(A)) \;
\Det {\cal M}_f \; \exp \{- S[\Chi ,A ] \, + \, \int d^{4}x \; j A \}
\en
\nopagebreak[4]
\be
S[\Chi ,A]\,=\, {1\over g^2}\int d^4x\, \left( {1\over
4}\Chi^a_{\mu\nu}\Chi^a_{\mu\nu} \pm {i\over 2}\Chi^a_{\mu\nu}
F^a_{\mu\nu}(A) \right)     \quad ,
\label{fiopi}
\en
If there were no gauge fixing the integral over the gauge field
would be Gaussian. At first sight it seems that with the
presence of the gauge fixing constraint the $A_{\mu }$-integration
can no longer be performed explicitly. However, one can transfer
the gauge fixing from the gauge potential $A_{\mu }$
to the field strengths. For this
purpose we insert the following identity into (\ref{fiopi})
\be
1 \, = \, \Det {\cal M}_g(\Chi )\int d(\theta ) \, \delta
( g^a(\Chi^\theta )) \quad ,
\en
where $ d(\theta ) $ is the invariant measure of the functional
integration over the group space, $\Chi^\theta $
denotes the gauge transformed of $\Chi $, and
$\Det {\cal M}_g(\Chi ) $ does not depend on $\theta $. The
key observation now is that the action $S[\Chi ,A] $ in (\ref{fiopi}) is
only invariant under simultaneous gauge transformations of $A$ and
$\Chi $. Therefore a change of the integration variable
$\Chi ^{\theta } \rightarrow \Chi $ implies also a change in the
gauge potential $A_{\mu } \rightarrow A_{\mu }^{(- \theta )}$ to leave
the exponent in (\ref{fiopi}) invariant. Because of the gauge
invariance of the measure and the determinants one then obtains
\be
Z \,=\, \int d(\theta )  \int {\cal D}\!\Chi {\cal D}A \; \delta
(f^a(A^{\theta^{-1}} )) \, \, \delta (g^b(\Chi )) \; \Det {\cal M}_f\,
\Det {\cal M}_g \;\; e^{- S[\Chi ,A]}  \quad .
\en
Now the integration over the gauge group can be performed again
yielding
\be
Z \,=\, \int {\cal D}\!\Chi {\cal D}A \; \delta
(g^b(\Chi)) \, \Det {\cal M}_g \;  \exp\left\{ - S[\Chi ,A]+\int d^4x \,
jA \right\}    \; .
\en
Fixing the gauge in terms of field strengths leaves a residual
invariance with respect to transformations, which leave the field
strengths  invariant. In the case of YM theories these transformations
belong to the discrete invariant subgroup of the gauge group.
Therefore (in contrast to the Abelian case) this residual invariance
is harmless.

Once the field strength $\Chi $ is gauge fixed there is no invariance
left in the potentials (up to the irrelevant residual invariance
mentioned above) and the
integration $\int {\cal D}A $ becomes unconstraint. For
non-singular $\hat\Chi^{ab}_{\mu\nu}\, $ it yields
\be
Z \,=\, \int {\cal D}\!\Chi  \; \delta ( g^a(\Chi )) \; \Det {\cal
M}_g \; ( \Det {i\over 2g} \hat\Chi )^{-1/2} \; \exp \{-
S_{FS}[\Chi ,j ] \}
\label{eq:2}
\en
\be
S_{FS}[\Chi,j] \,=\,  {1\over g^2}\int d^4x\, \left( {1\over
4}\Chi^a_{\mu\nu}\Chi^a_{\mu\nu} + {i\over 2}\Chi^a_{\mu\nu}
F^a_{\mu\nu}(J) +
j^{a}_{\mu } J^{a}_{\mu } - i \frac{g^{2}}{2} j^{a}_{\mu }
(\hat{\Chi }^{-1})^{ab}_{\mu \nu } j^{b}_{\nu }   \right)  \quad ,
\en
where
\be
J_\mu^a \,= \,\left(\hat\Chi^{-1}\right)^{ab}_{\mu\nu}
	   \partial_\rho \Chi^b_{\rho\nu}
\label{eq:4}
\en
is an induced gauge potential.
For singular $\hat \Chi $, integration over the gauge potential
$A^{a}_{\mu }(x)$ yields an expression similar to
(\ref{eq:2}), where the matrix $\hat \Chi $ is, however,
replaced by its projection onto the non-singular subspace.
But in addition constraints for the $\Chi $-integration result.
These constraints indicate that singular field configurations
$\hat \Chi $ are statistically suppressed.
Since $\Chi^{a}_{\mu \nu }$
behaves under gauge transformations as the field strength
$F^{a}_{\mu \nu } (A) $ the induced potential $J^{a}_{\mu } $
transforms precisely like the original gauge field $A^{a}_{\mu } $.
In practice, gauge fixing of the $\Chi $ can be done by using
the familiar gauges for the induced gauge potential
$J^{a}_{\mu } $ (\ref{eq:4}).

The presence of the  external source $j^{a}_{\mu } $
in the exponent of (\ref{eq:2}) ensures that Green's functions
of the original gauge potential $A^{a}_{\mu }$
are still accessible in the field strength formulation.

Finally in the field strength formulation a current current
interaction is induced which dominates the fermion dynamics
at low energies~\cite{re91,la91a,la91b,al92}.

\bigskip
\leftline{\it 4. Instantons in the field strength formulation \/ }
We are interested in a semiclassical analysis of the field strength
formulated YM theory. For simplicity we discard the external
gluon source $(j^{a}_{\mu }=0)$. The extrema of the action
$S_{FS}[\Chi ]= S_{FS}[\Chi ,j=0]$ occur for
\be
g \Chi^{a}_{\mu \nu } \; = \; -i \, F^{a}_{\mu \nu } (J) .
\label{eq:z}
\en
It was observed by Halpern~\cite{ha77} that the effective action of the
$\Chi $ field is extremized by the standard Polyakov t'Hooft
instantons. This fact is, however, not only true for the
standard SU(2) instanton but holds for any classical
solution extremizing the Yang-Mills action (\ref{eq:1a}).
For a proof we rewrite the classical YM equation of motion
(\ref{eq:3}) with the definition (\ref{eq:4}) as
\be
J^{a}_{\mu }( \, F(A) \, ) \; = \; A^{a}_{\mu }(x)
\label{eq:*}
\en
where we have for simplicity assumed that $\hat{F}^{ab}_{\mu \nu }(A) $
is not singular. Now let $A_{\mu }^{inst }$ denote
an instanton solution to (\ref{eq:*}). Since
$J^{a}_{\mu }( \Chi = -i F/g ) = J (F) $ it follows from (\ref{eq:*})
that the equation of motion of the field strength formulation
(\ref{eq:z}) is solved indeed for
\be
\Chi ^{a}_{\mu \nu } \; = \; - \frac{i}{g}
F^{a}_{\mu \nu } (A_{\mu }^{inst}) \; .
\label{eq:5}
\en
Furthermore, it follows then that also the classical action of
the instanton in the field strength formulation is the same as
in the standard approach
\be
S_{FS} [\Chi = - \frac{i}{g}
F(A_{\mu } ^{inst}) ] \; = \; S_{YM} [A^{inst}_{\mu }] \; .
\label{eq:y}
\en
The equivalence between both approaches holds, however,
not only at the classical level but also the quantum fluctuations
give identical contributions as we will
explicitly prove in the following for the
leading order in $\hbar $ corrections.

In the field strength formulation quantum fluctuations around a
background field were considered in \cite{re90a}. In the
semiclassical approximation the background field is chosen
as the instanton (\ref{eq:5}). If we expand the tensor field
$\Chi ^{a}_{\mu \nu }$ in terms of the t'Hooft symbols~\cite{tho76}
$\eta ^{i}_{\mu \nu }$, $\bar{\eta } ^{j}_{\mu \nu } $
\be
\Chi ^{a}_{\mu \nu } \; = \; \Chi^{a}_{i} Z ^{i}_{\mu \nu }
\; , \hbo
Z^{i}_{\mu \nu } \; = \; \{ \eta ^{i}_{\mu \nu }, \,
\bar{\eta }^{j}_{\mu \nu } \}
\label{eq:7a}
\en
the generating functional (\ref{eq:2}) becomes then
\be
Z[j=0] \; = \; Q_{FS} \;
( \Det \frac{i}{g} \hat{\Chi } ^{inst} )^{-1/2} \;
( \Det ' {\cal D}^{-1}_{FS} [ \Chi ^{inst} ] )^{-1/2} \;
 e^{ - S_{FS}[\Chi ^{inst} ] }
\label{eq:8}
\en
where we have used (\ref{eq:y}) and \cite{re90a}
\be
{\cal D}^{-1}_{FS} [\Chi ] (x_{1},x_{2}) ^{ab}_{ij} \; = \;
\frac{ \delta ^{2} S_{FS} [A] }{ \delta \Chi^{a}_{i} (x_{1})
\delta \Chi^{b}_{j} (x_{2}) } \mid _{\Chi ^{inst } }
\en
 is the second variation of the field strength action taken
at a background field $\Chi ^{inst}= -i F (A^{inst}) /g $.
The prime indicates again that zero modes are excluded. Their
contribution is included in the factor $Q_{FS}$.
For space-time dependent dependent background fields
$\Chi^{a}_{\mu \nu } = -i F^{a}_{\mu \nu }(x)/g $ one finds
\bea
{\cal D}^{-1}_{FS} [F ] (x_{1},x_{2}) ^{ab}_{ij} &=&
2 \delta ^{ab} \delta _{ij} \delta (x_{1},x_{2}) \; + \;
K^{ab}_{ij} [F] (x_{1},x_{2})
\label{eq:11} \\
K^{ab}_{ij} [F] &=& - \hat{D}^{ac}_{\kappa } (x_{1})
Z^{i}_{\kappa \mu } (\hat{F} ^{-1} )^{cd}_{\mu \nu }
\hat{D} ^{db}_{\lambda }(x_{1}) Z^{j}_{\lambda \nu }
\delta (x_{1},x_{2})
\label{eq:11a}
\ena
where
$\hat{D}^{ab}_{\mu }$ denotes here the covariant derivative
(\ref{eq:6}) with respect to the induced gauge potential
$J^{a}_{\mu }( \Chi = -i F ) = J^{a}_{\mu }(F)$
\be
\hat{D}^{ab}_{\mu }(x) \; = \; \partial _{\mu } \delta ^{ab}
\; - \; \hat{J}^{ab}_{\mu } (x) \; .
\label{eq:14}
\en
For a constant background field the above expressions for the
fluctuations reduce to the expressions given in \cite{re90a}.
Comparison of (\ref{eq:7}) and (\ref{eq:8}) shows if both
approaches give the same semiclassical result we should have
the relation
\be
Q \,
(\Det ' \,  g^{2} {\cal D} ^{-1}_{YM}[A_{inst}] (x_{1},x_{2})
)^{ - \frac{1}{2} } \; = \; C \, Q_{FS} \,
(\Det \hat{F}_{inst}  \;
\Det ' {\cal D}^{-1}_{FS} [F _{inst} ] )^{ - \frac{1}{2} } \; ,
\label{eq:9}
\en
where $C $ is an irrelevant (but non-vanishing )
constant, which does not depend on the
instanton solution.
We will now explicitly prove this relation.

\bigskip
\leftline{\it 5. Equivalence proof \/ }
In order to establish the validity of (\ref{eq:9})
we cast the functional matrix (\ref{eq:10}) of the standard approach YM
into the form of its field strength formulated
counterpart (\ref{eq:11}) by writing
\be
2 g^{2} {\cal D} ^{-1}_{YM} [F] (x_{1},x_{2}) ^{ab}_{\mu \nu } \; = \;
(\hat{F}^{1/2}) ^{ac}_{\mu \sigma }(x_{1}) \, M^{ce}_{\sigma \rho }
[F] (x_{1},x_{2}) \, (\hat{F}^{1/2}) ^{eb}_{\rho \nu } (x_{2})
\label{eq:Ma}
\en
\be
M^{ab}_{\mu \nu } [F] = 2 \delta ^{ab} \delta _{\mu \nu }
\delta (x_{1},x_{2}) \; + \; (\hat{F}^{-1/2} ) ^{ad}_{\mu \sigma } \,
\int d^{4}x \;
\frac{ \delta F^{c}_{\kappa \lambda }(x) }{ \delta A^{d}_{\sigma }(x_{1}) }
\frac{ \delta F^{c}_{\kappa \lambda }(x) }{ \delta A^{e}_{\rho }(x_{2}) } \;
(\hat{F}^{-1/2} ) ^{eb}_{\rho \nu } \, .
\label{eq:M}
\en
We first prove that $M[F]$ has the same eigenvalues
(including the zero modes )
as ${\cal D}^{-1}_{FS} [F]$. Let $\phi ^{a}_{\mu }(x)$
and $\lambda $ denote the eigenvectors and eigenvalues of $M$:
\be
\int d^{4}x_{2} \; M^{ab}_{\mu \nu } [F] (x_{1},x_{2}) \, \phi ^{b}_{\nu }
(x_{2}) \; = \; \lambda \, \phi ^{a}_{\mu } (x_{1})
\en
Defining
\be
\psi ^{c}_{\kappa \lambda } (x) \; := \;
\int d^{4}x_{2} \;
\frac{ \delta F^{c}_{\kappa \lambda }(x) }{ \delta A^{e}_{\rho }(x_{2}) } \;
(\hat{F}^{-1/2} )^{eb}_{\rho \nu } (x_{2}) \,
\phi ^{b}_{\nu } (x_{2})
\label{eq:yy}
\en
the eigenvalue equation becomes
\be
\int d^{4}x \; \bar{K} ^{dc}_{\rho \sigma , \kappa \lambda }
[F] (y,x) \, \psi ^{c}_{\kappa \lambda } (x) \; = \;
(\lambda -2) \psi^{d}_{\rho \sigma }(y)
\label{eq:12}
\en
where
\be
\bar{K} ^{dc}_{\rho \sigma , \kappa \lambda }
[F] (y,x) \; = \; \int d^{4}x_{1} \;
\frac{ \delta F^{d}_{\rho \sigma }(y) }{ \delta A^{a}_{\mu }(x_{1}) } \,
[ \hat{F} ^{-1} (x_{1}) ]^{ab}_{\mu \nu }
\frac{ \delta F^{c}_{\kappa \lambda }(x) }{ \delta A^{b}_{\nu }(x_{1}) } \; .
\label{eq:zz}
\en
By construction the amplitudes $\psi ^{a}_{\mu \nu }$ are
antisymmetric in $(\mu ,\nu )$ and can hence be expanded in terms
of t'Hooft symbols (c.f.\ (\ref{eq:7a}))
\be
\psi ^{a}_{\mu \nu } \; = \; \psi ^{a}_{i} Z^{i}_{\mu \nu } \; .
\en
The eigenvalue equation (\ref{eq:12}) then reads
\bea
\int d^{4}x \; \bar{K} ^{dc}_{i j }
[F] (y,x) \, \psi ^{c}_{j } (x) & = &
(\lambda -2) \psi^{d}_{i }(y)  \\
\bar{K} ^{ab}_{ij}(y,x)  &=& \frac{1}{4} Z^{i}_{\mu \nu }
\bar{K} [F](y,x)^{ab}_{\mu \nu , \kappa \lambda }
Z^{j}_{\kappa \lambda } \; .
\ena
Inserting  the explicit form of
$\delta F^{a}_{\kappa \lambda }(x)  / \delta A^{b}_{\mu }(y) $
(\ref{eq:13}) into $\bar{K} [F] $ (\ref{eq:zz}) the integration over
the intermediate coordinate
$x_{1}$ can be carried out upon using \hfil \break
$\hat{D}^{ab}_{\mu } (x) \delta (x-y) \; = \; - \hat{D}^{ba}_{\mu } (y)
\delta (x-y)$. Exploiting also the antisymmetry of the $Z^{i}_{\mu \nu }$
the kernel $\bar{K}[F]$ takes the form
\be
\bar{K} ^{ab}_{ij} (x,y) \; = \; - \hat{D} ^{ac}_{\mu }(x)
Z^{i}_{\mu \kappa } ( \hat{F}^{-1}(x) )^{cd}_{\kappa \lambda }
\hat{D}^{db}_{\nu } (x) Z^{j}_{\nu \lambda } \, \delta (x-y) \; .
\label{eq:12a}
\en
For an instanton background field $A_{\mu }$ we have in view of
(\ref{eq:*}) $J^{a}_{\mu } (x) = A^{a}_{\mu }(x) $ and
the covariant derivatives in (\ref{eq:6}) and (\ref{eq:14})
are the same so that the kernels $K$ (\ref{eq:11a})
and $\bar{K}$ (\ref{eq:12a}) are identical.
We thus proved that all eigenvalues of $M$ (\ref{eq:M})
are also eigenvalues of ${\cal D}^{-1}_{FS}$, and the
corresponding eigenvectors are related by (\ref{eq:yy}).

On the other hand, not
all eigenvalues of  ${\cal D}^{-1}_{FS}$ (\ref{eq:11}) are also eigenvalues
of $M$ (\ref{eq:M}). This is because the transformation (\ref{eq:yy})
maps the Hilbert space of eigenstates of $M$ of
dimension $n=D(N^{2}-1)$ onto a subspace of the
$ m = \left( \begin{array}{c} D \\ 2 \\ \end{array} \right) (N^{2}-1) $
dimensional space of eigenvectors of $\bar{K}$.
This implies that the $m-n$ additional eigenvectors of $\bar{K}$, denoted
by $\psi ^{(0) \; c }_{\mu \nu }$, are zero modes
\be
\int d^{4}x \; \bar{K} ^{dc}_{\rho \sigma , \kappa \lambda }
(y,x) \, \psi ^{(0) \; c}_{\kappa \lambda } (x) \; = \; 0
\en
satisfying
\be
\int d^{4}x_{1} \;
\frac{ \delta F^{c}_{\kappa \lambda }(x) }{ A^{a}_{\mu }(x_{1}) }
\psi ^{(0) \; c }_{ \kappa \lambda } (x_{1}) \; = \; 0 \; .
\en
These additional zero eigenvalues of $\bar{K}$ give rise to additional
eigenvalues $2$ of ${\cal D}^{-1}_{FS}$ in the field
strength formulation. The latter contribute only an
irrelevant constant to the functional determinant of ${\cal D}^{-1}_{FS}$,
which can be absorbed into the constant $C$ in (\ref{eq:9}).

If there were no zero modes the proof of (\ref{eq:9}) would be completed.
This is because in the absence of zero modes $(Q=Q_{FS}=1)$
from (\ref{eq:Ma}) would follow
\be
\Det {\cal D}^{-1}_{YM} \; = \; \Det \hat{F} \; \Det M
\en
and we have shown above that
\be
\Det ' M[F] \; = C \, \Det ' {\cal D} ^{-1}_{FS}
\en
provided the same regularization is used.

In the presence of zero modes some more care is required.
Their contribution~\cite{ber79} to the functional integrals is
represented by the preexponential factors in (\ref{eq:7})
and (\ref{eq:8})
\be
Q \; = \; \Det ^{ 1/2 } \langle \delta _{i} A_{cl} \mid \delta_{k}
A_{cl} \rangle
\; \hbox to 2cm {\hfil and \hfil } \;
Q_{FS} \; = \; \Det ^{ 1/2 } \langle \delta _{i} \Chi_{cl} \mid
\delta _{k} \Chi_{cl} \rangle
 \label{eq:100}
\en
respectively. Here $\delta _{i} A_{cl} $ denotes the variation
of the classical (instanton) solution with respect to its
$i$th symmetry parameter, which  is the (unnormalized) zero mode.
Its counterpart $\delta _{i} \Chi_{cl} $  in the field strength
formulation is related to $\delta _{i} A_{cl}$ by
\be
\delta _{i} (\Chi_{cl}\,)^{a}_{\kappa \lambda } (x) \; = \;
- \frac{i}{g} \delta _{i} F^{a}_{\kappa \lambda } [A_{cl} ]
\; = \; - \frac{i}{g} \int d^{4}x \;
\frac{ \delta F^{a}_{\kappa \lambda } }{ \delta A^{b}_{\nu } (x) }
\delta _{i} {A_{cl}\,}_\nu^b (x) \; .
\en
Exploiting the fact ${\cal D}^{-1}_{YM}\, \delta _{i} A_{cl} = 0$
one readily verifies that
\be
Q_{FS}^{2} \; = \; \Det \langle \delta _{i} A_{cl} \mid \hat{F} \mid
\delta _{k} A_{cl} \rangle \; = \; Q^{2} \; \Det ^{(0)} \hat{F} \; ,
\label{eq:101}
\en
where $\Det ^{(0)} \hat{F}$ is the determinant of the matrix arising
from projection $\hat{F}$ onto the space of zero modes of
${\cal D}^{-1}_{YM}$.
Inserting (\ref{eq:101}) into (\ref{eq:9}) it remains to be proven that
\be
\frac{1}{ \Det ' {\cal D}^{-1}_{YM} } \; = \;
\frac{ \Det ^{(0)} \hat{F} }{ \Det \hat{F} \; \Det ' {\cal D}^{-1}_{FS} }
\; . \label{eq:102}
\en
In order to extract $\Det' {\cal D}_{FS}^{-1}$ from
$\Det' {\cal D}_{YM}^{-1}$ it
is convenient to introduce the complete set of orthonormal eigenvectors
$\varphi _{i}$ and $\phi_i$ of the symmetric matrices
${\cal D}^{-1}_{YM} = \hat{F}^{1/2} M \hat{F}^{1/2}$ and M.
We denote the normalized zero and non-zero modes of
${\cal D}_{YM}^{-1}$ $(M)$ by $\varphi ^{(0)}_{i}$,
$(\phi ^{(0)}_{i})$ and $\varphi '_{i}$, $(\phi '_{i})$,
respectively.
Accordingly $\Det ^{(0)} \hat{F}$ and $\Det ' \hat{F}$ denote the
respective subspace determinants of $\hat F$.
{}From the defining equation (\ref{eq:Ma})  follows that the vectors
$\lbrace \hat F^{1/2} \varphi^{(0)}_k \rbrace $ span the space of zero
modes $\lbrace \phi_k^{(0)} \rbrace $ of $M$ and conversely, the
$\lbrace \hat F^{-1/2} \phi^{(0)}_k \rbrace $ span the space of zero
modes $\lbrace \varphi_k^{(0)} \rbrace $ of $\Det' {\cal
D}_{YM}^{-1}$. From the orthogonality of the $\phi_k $ and the
$\varphi_k$ then follows
\be
\phi '^{T}_{i} \hat{F}^{1/2} \varphi^{(0)}_{k} \; = \; 0
\hbox to 2cm {\hfil and \hfil }
\varphi '^{T}_{i} \hat{F}^{-1/2} \phi^{(0)}_{k} \; = \; 0
\label{eq:46}
\en
respectively, implying that we may write
\be
\hat{F}^{1/2} \varphi ^{(0)}_{l} \; = \; U_{lm} \phi ^{(0)}_{m}
\hbox to 2cm {\hfil and \hfil }
\hat{F}^{-1/2} \varphi '_{l} \; = \; O_{lm} \phi '_{m} \; .
\label{eq:la}
\en
For later use we calculate the determinant of the matrices $U$ and $O$.
Due to the orthonormality of the eigenvectors $\varphi _{l} $
we have
\be
\varphi ^{(0) \, T}_{i} \varphi ^{(0)} _{k} \; = \; U_{il} U_{km} \,
\phi^{(0) \,T}_{l} \hat{F}^{-1} \phi^{(0)}_{m} \; = \; \delta _{ik}  \; ,
\qquad
\varphi '^{T}_{i} \varphi '_{k} \; = \; O_{il} O_{km} \,
\phi '^{T}_{l} \hat{F} \phi '_{m} \; = \; \delta _{ik}
\en
and thus
\be
1 \; = \; \Det U \, \Det^{(0)} \hat{F}^{-1} \, \Det U^{T}
\hbox to 2cm {\hfil and \hfil }
1 \; = \; \Det O \, \Det ' \hat{F} \, \Det O^{T}  \quad ,
\en
implying
\be
\Det ' \hat{F} \; = \; ( \Det O )^{-2} \; ,
\hbox to 2cm {\hfil and \hfil }
\Det ^{(0)} \hat{F} \; = \; (\Det U )^{2}  \; .
\label{eq:lb}
\en
In order to show that
\be
\Det \hat{F} \; = \; \Det ' \hat{F} \; \Det ^{(0)} \hat{F} \; ,
\en
we note that the orthonormal sets $\varphi _{i}$ and $\phi _{k}$
are related by an orthogonal transformation. Therefore
we may write
\be
\Det \hat F^{1/2} \, =\,  \Det \left( \varphi '^T_l , {\varphi
^{(0)}}^T_l \right) \hat F^{1/2} \left( \begin{array}{l} \phi '_k
\\ \phi^{(0)}_k \end{array}  \right)  \quad .
\en
In view of (\ref{eq:46}) the matrix on the right hand side
is of triangular block form.
\bea
\Det \hat F^{1/2} &= &\Det\left( \varphi '^T_l \hat F^{1/2} \phi
'_k\right) \; \Det \left(\varphi^{(0)\, T}_l \hat F^{1/2}
\phi^{(0)}_k\right)    \cr
	&= &\Det\left( O_{lm} \phi '_m \hat F \phi '_k \right) \;
\Det \left( U_{lm} \phi^{(0)\, T}_m \phi^{(0)}_k \right) \cr
        &= &\Det O \; \Det '\hat{F} \; \Det U \cr
	&= &\left(\Det ' \hat F\right)^{1/2} \; \left( \Det^{(0)}
\hat F\right)^{1/2}    \quad ,
\ena

An analogous manipulation, again using (\ref{eq:46}) and
(\ref{eq:la}), shows that the determinant $\Det ' {\cal
D}^{-1}_{YM} $  in (\ref{eq:102}) factorizes
\bea  \Det ' {\cal D}^{-1}_{YM } &= &\Det\left( \varphi '^T_k \hat
F^{1/2} \phi '_m \phi '^T_m M \phi '_n \phi '^T_n \hat
F^{1/2} \varphi '_l \right) \cr
	&= &\Det O \; \Det ' \hat F \;\Det ' M\; \Det '\hat F\;\Det
O^T \, = \, \Det ' \hat F \; \Det ' M \quad . \ena

This completes the proof of (\ref{eq:102}), which establishes
explicitly the equivalence of the field strength
formulation and the standard formulation at the
semiclassical level.

In the so called field strength approach of \cite{sch90}
the $(\Det \hat{\Chi } )^{-1/2}$ arising from the integration over the
gauge field is included into an effective action
\be
S_{FSA} \; = \; \frac{1}{4} \int d^{4}x \; \{
\Chi^{2} + \frac{ \mu ^{4} }{2} \tr \ln \frac{i}{g} \hat{\Chi } / \mu ^{2}
+ \frac{i}{2g} \Chi F(J) \; \}
\label{eq:15}
\en
where the scale $\mu $ arises from the regularizations of
$ \Tr \ln \frac{i}{g} \hat{\Chi} $. Due to the appearance of the
scale this effective action does no longer tolerate
instantons as stationary points~\cite{la92b} as one might have
expected since instantons have a free
scale (size) parameter.
This was explicitly shown already in~\cite{sch90} for t'Hooft-Polyakov
instantons.
Instead of instantons
the effective action (\ref{eq:15})
has (up to gauge transformations) constant solutions
$\Chi = -i G$,
which can be interpreted as instanton solids~\cite{la92b}.
If one considers fluctuations around these constant solutions
the propagator of the fluctuations is given by \cite{re90a}
\be
{\cal D} ^{-1}_{FSA} \; = \; 2 \, ( 1 \, + \, C) \, + \, K
\label{eq:16}
\en
where $K$ is defined by (\ref{eq:11a}) and the extra term
\be
C^{ab}_{ij} \; = \; \frac{1}{2} \tr [
\hat{G}^{-1} T^{a} Z^{i} \hat{G}^{-1} T^{b} Z^{j} ] \; \hbo
(T^{a})^{bc} = f^{abc}
\en
arises from the $ \Tr \ln \frac{i}{g} \hat{\Chi} $ term in the effective
action (\ref{eq:15}).
For a constant background field the additional zero modes
$\psi ^{(0)}$ of $\bar{K}$ found above in the instanton background
correspond to non-propagating
(non-dynamical) modes in the  field strength formulation.
This also reflects the fact that the field
strength formulation, although using the larger number of field
variables $\Chi^{a}_{\mu \nu }$, contains the same number
of propagating (dynamical) modes as the standard formulation
as was already observed in \cite{re90a}.

For large momenta $p^{2}$ the term $C^{ab}_{ij}$ is
however negligible compared to the $p$-dependent term $K(p)$ and
the propagator of the fluctuations in the field strength approach
${\cal D}_{FSA}$ (\ref{eq:16}) reduces to ${\cal D} _{FS} $
(\ref{eq:11}), which we have shown to yield the same quantum
effects as the standard propagator. This implies that the FSA yields
the same asymptotic $(p^{2} \rightarrow \infty )$ gluon propagator
$( \sim  1/p^{2} )$ as the standard formulation as will be explicitly
demonstrated elsewhere~\cite{lo92}.
%
\medskip

\end{document}